# A natural-language-based approach to intelligent data retrieval and representation for cloud BIM

Jia-Rui Lin, Zhen-Zhong Hu*, Jian-Ping Zhang, and Fang-Qiang Yu

*Department of Civil Engineering (Tsinghua University), Haidian District, Beijing 100084, PR China*

**Abstract:** As *the information from diverse disciplines continues to integrate during the whole life cycle of an Architecture, Engineering, and Construction (AEC) project, the BIM (Building Information Model/Modeling) becomes increasingly large. This condition will cause users difficulty in acquiring the information they truly desire on a mobile device with limited space for interaction. The situation will be even worse for personnel without extensive knowledge of Industry Foundation Classes (IFC) or for non-experts of the BIM software. To improve the value of the big data of BIM, an approach to intelligent data retrieval and representation for cloud BIM applications based on natural language processing was proposed. First, strategies for data storage and query acceleration based on the popular cloud-based database were explored to handle the large amount of BIM data. Then, the concepts "keyword" and "constraint" were proposed to capture the key objects and their specifications in a natural-language-based sentence that expresses the requirements of the user. Keywords and constraints can be mapped to IFC entities or properties through the International Framework for Dictionaries (IFD). The relationship between the user's requirement and the IFC-based data model was established by path finding in a graph generated from the IFC schema, enabling data retrieval and analysis. Finally, the analyzed and summarized results of BIM data were represented based on the structure of the retrieved data. A prototype application was developed to validate the proposed approach on the data collected during the construction of the terminal of Kunming Airport, the largest single building in China. The case study illustrated the following: (1) relationships between the user requirements and the data users concerned are established, (2) user-concerned data can be automatically retrieved and aggregated based on the cloud for BIM, and (3) the data are represented in a proper form for a visual view and a comprehensive report. With this approach, users can significantly benefit from requesting for information and the value of BIM will be enhanced.*

*To whom correspondence should be addressed. E-mail: huzhenzhong@tsinghua.edu.cn.

## 1 INTRODUCTION

Building information model/modeling (BIM) represents a new paradigm for the Architecture, Engineering, and Construction (AEC) industry, encouraging the integration of the roles of all stakeholders in a project (Azhar, 2011). Based on reports from the McGraw-Hill Construction and Pike Research, the number of industry participants, construction projects and related companies that are adopting BIM for technical support is increasing. Research indicates that the BIM adoption rate among AEC companies will continue to increase (Jennifer, 2012).

Efforts have been exerted to promote the creation, sharing and integration of BIM as well as collaboration and Knowledge Management (KM) throughout the life cycle of an AEC project. Two different methods were adopted for interoperability and collaboration based on BIM:

1. Methods based on ontologies and semantic web: Anumba (et al., 2008) concluded that explicit domain ontologies play an important role in supporting collaboration and decision making in various aspects of design and construction by generating a common representation of the problem-solving domain for the various disciplines. With this idea, ontologies and information exchange should be first studied and existing research shows some breakthroughs. Existing ontologies were reviewed, and suggestions for linking virtual models and construction components based on ontologies were proposed (Sørensen et al., 2010) for information sharing in construction and building operation management. An approach to achieving interoperability between Web-based construction product catalogues was also presented (Kong et al., 2005).

2. Methods based on Industry Foundation Classes (IFC): IFC is a comprehensive representation of the building model and the rules and protocols that define building data proposed by the International Alliance of Interoperability (IAI, also known as BuildingSMART), to address the inadequate interoperability problem (Laakso et al., 2012). A theoretical framework of technical requirements for using an IFC-based BIM-server as a multidisciplinary collaboration platform was also discussed (Singh et al., 2011). Procedures for developing Information Delivery Manual (IDM) were defined (Eastman et al., 2009), to facilitate precise data exchange based on IFC. A mobile model based on IFC and 4D model was also presented for bridge management (Hammad et al., 2006).

As proposed approaches and tools for interoperability and collaboration increase, more stakeholders will be involved, more information will be integrated into the BIM, and the size of the data will continue to increase. This situation will cause users difficulty in acquiring the information they truly desire. Even experts of the BIM software or developers are required to consider the deluge of information (Jasper, 2011) by using several manipulations. With the continued increase in data size and functions of the BIM software, additional time is needed to study the software, and users will have more difficulty in properly obtaining useful but concentrated information. Automated processing and extraction of information would avoid the tediousness and error prone of manually work (Boukamp and Akinci, 2007). In addition, it is getting more popular to retrieve information from the web using mobile terminals, whose display size is much smaller than conventional laptops and desktops, requiring a simple and convenient interaction interface with limited spaces on the devices. Voice assistants such as Siri, Google Now and Cortana have set a new example for request for information. Hence, flexible and effective data acquisition from a huge BIM is required for both experts and non-experts of the BIM software, and feasible data visualization or representation methods will strengthen the value of BIM.

This paper presents a cloud-based framework for intelligent BIM data retrieval and representation. First, a cloud-based database was established for BIM data storage and manipulation. Then, natural language processing (NLP) was adopted to process user input in natural language, that is, keywords were extracted and then mapped to IFC entities or attributes. This process laid the basis for retrieving information in an IFC-structured BIM data model. Finally, the retrieved data results were analyzed and represented in the form of charts, tables, animations, or a combination of forms based on the expectations of users.

In Section 2, related research is reviewed and the possibility of intelligent BIM data retrieval and representation for BIM applications is discussed. Section 3 proposes principles for designing a cloud-based data storage and implementation of the database. Section 4 describes detailed processes of keyword extraction based on NLP and its mapping method to IFC entities supported by the International Framework for Dictionaries (IFD). Section 5 illustrates how information is retrieved, analyzed and represented. Section 6 shows an actual case study of Kunming Airport, the largest single building in China, to validate the proposed approach. Finally, Section 7 discusses the conclusions and future works.

## 2 LITERATURE REVIEW

### 2.1 Related works
#### 1. Interoperability based on IFC and IFD

The National Building Information Modeling Standards Committee of the United States explains that BIM encompasses information throughout the life cycle of a facility, supporting multidisciplinary collaboration and decision making (NBIMS, 2012). IFC is a data protocol proposed for sharing information between different software. Backed by years of development and improvement, the IFC standard is widely accepted for information management. An IFC-based system (Lee et al., 2003) was proposed for design information management, and some IFC-based BIM servers (Beetz et al., 2010; Jørgensen et al., 2008; Kang and Lee, 2009) were also developed for information sharing, extraction and integration. Based on BIMserver.org (Beetz et al., 2010), an open query language for BIM called BIMQL was developed (Mazairac and Beetz, 2013), providing flexible data retrieval interface that is domain specific and platform independent. An open repository for IFC model analysis to facilitate the interoperability of building information was also presented (Amor and Dimyadi, 2010). Meanwhile, experiences with issues of model-based interoperability in exchanging BIM between various tools were reported (Steel et al., 2012). Research on extending IFC by eXtensible Markup Language (XML), and information exchange with the help of model view (Eastman et al., 2009; Fu et al., 2006) or IDM (Zhang et al., 2012; Kim et al., 2010) were also discussed. Therefore, IFC, with feasible extending mechanism and numerous related tools, lays a solid foundation for BIM interoperability.

However, as stated in the IFD white paper (BuildingSMART, 2008), a controlled vocabulary of construction terminology, which IFC lacks, is essential to support data exchange. With the support of multi-language terminology, IFD provides a mapping method from concepts to IFC entities and attributes (Zhang et al., 2012), supporting distinguishing concepts from specific linguistic instances. Shayeganfar (et al., 2008) conducted a case study on how to implement an IFD library using semantic web technologies, which bridge the gap between BIMs and web services. Thus, with the support of IFD library and semantic web, terminology or its synonym in a specific language can be mapped to the entity in a data schema like IFC.

#### 2. Cloud and BIM

Using and sharing the information in BIM based on a stand-alone system become more difficult as the BIM becomes bigger, cloud computing will help in information manipulation (Chuang et al., 2011). Investigation conducted by Redmond (Redmond et al., 2012) shows that some enterprises are already engaged in developing a cloud BIM system, and the cloud will enhance the information sharing and visualization. With a cloud BIM: (1) the possibility of many disciplines collaborating on the same system will be enhanced (Redmond et al., 2012); (2) the use of the web service will bring an easy and flexible access to information and resources; and (3) handling a large amount of data using an extensible platform will be possible.

Nowadays, BIM 360 Glue, BIM+, BIMcloud and other products based on cloud BIM are already released. However, these products mainly focus on data sharing, collaboration and visualization. For the lack of information about how the data are stored and accessed, it is hard to extend the products in data storage, manipulation and visualization. Since building a cloud computing platform that utilizes a non-relational, distributed database like MongoDB is now popular (Membrey et al., 2010), it is valuable and feasible to explore how IFC models can be stored and shared based on these NoSQL databases.

### 3. Data retrieval and visualization

To facilitate the acquisition of the information, domain knowledge was utilized to assist the information retrieval of an online product (Lin and Soibelman, 2009). A question answering system (Cheng et al., 2002) for project management and system for context sensitive information in building performance management (Keller et al., 2008) were proposed to enhance the data retrieval. To facilitate the data retrieval for BIM, different approaches such as BIMQL (Mazairac and Beetz, 2013) and graph-based data retrieval (Langenhan et al., 2013) were proposed. However, the former provides a query language similar to SQL, which most of the users are unfamiliar with. Using a SQL-like query language is difficult for a non-expert. The latter approach is designed to deal with spatial information. Both approaches did not provide a feasible data representation interface.

Data representation and visualization play an important role in clearly and effectively expressing information through different methods. Studies on visualization of categorical data (Chang and Ding, 2005), time oriented data (Klimov et al., 2010) and so on were discussed. A graphical representation was presented, supporting multiple design representation subsystems and their interactions to facilitate design collaboration in 3D virtual worlds (Gu and Tsai, 2010). An approach to modeling the degradation visualization of building flooring systems was proposed for building maintenance (Khosrowshahi et al., 2014). Construction quality was improved and coordination was enhanced among trade contractors based on a visual process for planning construction (Tan et al., 2005).

### 2.2 Discussion

IFC, the object-oriented and semantic schema, provides a solid foundation for BIM interoperability. Most BIM applications now adopt IFC as a data exchange standard, a large amount of data is accumulated and will continue to increase. It is of potential to carry out a data retrieval and analysis service based on the IFC-structure data model. Meanwhile, related studies show that implementing a cloud BIM is possible and will improve the collaboration and interoperability based on BIM. And an intelligent data retrieval approach based on natural language will provide a flexible way to retrieve BIM data for a wide range of stakeholders. A data representation method will enhance the value of the data retrieval approach.

Natural Language Processing (NLP) is widely used and deeply explored as a computerized approach to analyze text based on both a set of theories and a set of technologies (Liddy, 2001). Tools including Stanford parser (De Marneffe et al., 2006), and NLTK (Bird et al., 2009) were already developed to process text in different languages.

In conclusion, an urgent demand and a good foundation exist for intelligent BIM data retrieval and representation based on the requirements of non-experts. With a large amount of BIM data collected from different BIM applications, further promoting the value of cloud BIM based on flexible retrieval and feasible representation is important.

### 3 CLOUD-BASED DATA STORAGE

With a cloud, people of diverse disciplines in different places can access the same data and effectively collaborate and communicate. The cloud-based BIM data storage is adopted to flexibly manipulate the huge BIM and to facilitate information retrieval for different users.

Figure 1 shows that the cloud consists of a number of clusters that provide information and data manipulation functions for the owners, contractors and other clients. Each cluster of the cloud consists of metadata that defines what types of information are stored in this cluster and a group of NoSQL databases that persist data in an IFC format BIM. A MapReduce framework, which is a programming model and an associated implementation that parallelizes data processing across large-scale clusters of machines (Dean and Ghemawat, 2008), is usually provided to implement the cloud. The framework is divided into two parts: map, and reduce. The former is a function that parcels out work to different nodes in the distributed cluster, and the latter is another function that collates the work and resolves the results into a single value. Based on the MapReduce framework of the cloud platform, analysis and manipulation will be enhanced for fast access to information.

An interface for the column oriented database was implemented in the BIM server (Beetz et al., 2010), it is possible to switch to other NoSQL databases and other types of databases (such as relational database) by implementing the BIM database interface. However, the BIM server does not provide a proper interface to integrate the MapReduce framework or similar function provided by distributed databases, such as HBase or MongoDB, a new database is necessary for these requirements.

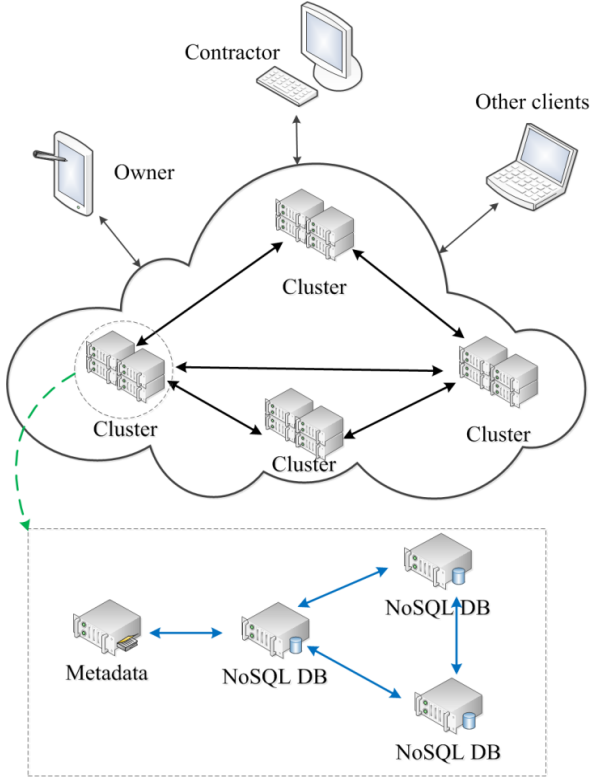

**Figure 1** Framework of cloud-based data storage

After comparing different databases, a popular free document database called MongoDB was selected as the foundation of the cloud-based database in this study, because of the following advantages.
(1) Scalability: MongoDB is a NoSQL database with built-in features, such as sharding and replication. This advantage allows the database to be scaled to as many servers as possible, and to automatically distribute and synchronize data among them.
(2) MapReduce framework: MongoDB provides a lightweight MapReduce framework based on javascript. It is convenient to implement customized MapReduce operations.
(3) File storage: With the GridFS of MongoDB, large files such as attached videos and pictures for different products can be stored with metadata.
(4) Full index support and rich, flexible query: In MongoDB, users can create indices for almost any field of an object. This feature will allow a fast query of the data in the database. MongoDB also provides a flexible interface for querying data.
(5) Other: With drivers for different programming languages and a configurable serialization mechanism, MongoDB supports a wide range of developers. Thus, it is easy to control the serialization of an object.

### 3.1 Serialization strategies for IFC-based BIM data

In MongoDB, objects of the same type are stored in a collection and each object is serialized as a bson (similar to json) document in the collection. The attribute of an object can be serialized as a single value if it is a simple type, such as string or integer. While for a complex type, it should be serialized as a nested sub-document or as a document in another collection (Figure 2).

For an object-oriented data model, such as IFC, objects in the model are referenced to one another, which in some situation causes self-reference, resulting in problems when serializing the entities into a database. With a deep inherit chain, serializing an object of the IFC data model to a MongoDB document may also result in a document with deep nested sub-documents. It will lead to inefficient data retrieval and analysis. Given that IFC is designed for data exchange, the geometric information of an IFC data model is too complicated and inefficient for display in a 3D view. Thus, the geometric information should be converted into a display friendly format and stored separately.

Therefore, different serialization strategies should be adopted when designing the IFC-based database on MongoDB. Entities in an IFC data model can be classified into five parts:

$$M = \bigcup\{O, RL, P, G, RLx\} \tag{1}$$

where
(1) $O$ is a set of entities that inherit from IfcObjectDefinition. All entities in $O$ should be persisted as documents (Figure 2) in different collections in MongoDB. If an attribute of an entity is in set $O$, then the ID of this attribute should be saved instead of its data.
(2) $RL$ are the entities that inherit from IfcRelationship. For entities in $RL$, the same strategy as (1) should be used.
(3) $P$ represents data types and entities defined in the resource layer of the IFC schema. All entities in set $P$ can be considered as embedded documents (similar to OwnerHistory and OwningUser in Figure 2) of MongoDB.
(4) $G$ is the set of entities for geometric representation. An entity $g$ in $G$ that contains the geometric information can be saved as a sub-document. However, this condition will result in a weight document, making it inefficient for data query. Thus, these data should be saved in separate collections, and the IDs of the data should be kept in the document of the entity that referenced them (as RepresentationId in Figure 2). To

efficiently represent geometric data, these data should be converted to a display-friendly format, such as wavefront obj or collada in advance, and the data should be cached into the GridFS of MongoDB. This strategy is also adopted because few changes were made to the shapes and dimensions of a building element in this research.

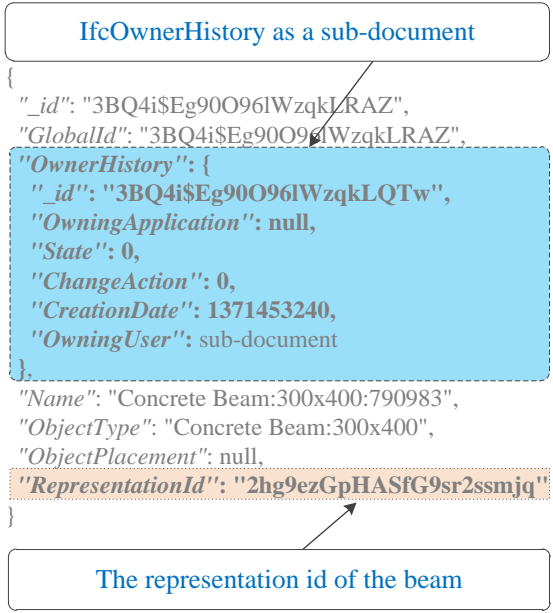

**Figure 2** Example of serializing an instance of IfcBeam in MongoDB

(5) $RLx$ are some entities in $O$ or $P$ but function as relational objects. An example of these entities is IfcMappedItem, which is the use of an existing representation as a representation item for another representation. Entities in $RLx$ can be serialized as sub-documents or as documents in different collections. The former is suitable for query while the latter is feasible for update. Which entity is contained in $RLx$ should be determined by the developers according to different user case.

Basic insert, query and update tests on a single machine showed that MongoDB is four times faster in data insertion and six times faster in data updating than SQL Server, whereas the query speed of MongoDB is about half of that of SQL Server. Similar results were also provided in other speed tests (Bromberg, 2013). Following the above-mentioned strategies, the IFC-based database built on MongoDB is a semi-structure database. Most of the properties of an entity will be embedded in the main document of the entity, it is flexible and fast to access the properties of an object in the model. Thus, MongoDB is a feasible storage for the BIM.

### 3.2 MapReduce-based pre-join for query acceleration

As previously mentioned, entities in an IFC model are stored in different collections, while the data of the same collection is distributed to different clusters. Since MongoDB doesn't provide any operation similar to join operation in relational database, if entity B is referenced by entity A while they are stored in different collections, then two query calls are necessary to obtain both of them. To eliminate this type of call at the client side, a pre-join approach based on the MapReduce framework was adopted. An illustration on how to join data in collection B to collection A is shown in Figure 3 and explained as the follows.

(1) For each entity $entA_i = \{AKey_i, P_{ai}, BKey_i\}$ in collection $A$, temporary entity $entTemp_{ai}$ is first created through the map function as follows:
$$entTemp_{ai} = \{AKey_{ai}, P_{ai}, P_{b1i}, P_{b2i}, BKey_{bi}\} \quad (2)$$
While in this temporary collection $P_{b1i}, P_{b2i}, BKey_{bi}$ are considered *null*;

(2) Similarly, for entity $entB_j = \{BKey_j, P_{b1j}, P_{b2j}\}$ in collection $B$, $entTemp_{bj}$ is created through the map function as follows:
$$entTemp_{bj} = \{AKey_{aj}, P_{aj}, P_{b1j}, P_{b2j}, BKey_{bj}\} \quad (3)$$
While $AKey_{aj}, P_{aj}$ are pre-defined as *null*;

(3) Then, a reduce function is executed to merge a temporary entity with the same $BKey_{tk}$. When more than one $AKey_{tk}$ occurs, all data from collection A (in the form of $\{AKey_{ai}, P_{ai}\}$) should be stored as an array of sub-documents (part with a blue background in Figure 3). The temporary entity should be similar to the following:
$$entT_k = \{BKey_{bk}, P_{b1k}, P_{b2k}, Col\} \quad (4)$$
where $Col$ is a collection of
$$entCol = \{AKey_{ai}, P_{ai}\} \quad (5)$$

(4) Finally, for each $entT_k$ in the temporary collection and for each $entCol$ in $Col$ of $entT_k$, a document in the form of equation (2) should be created.

$(i, j, k = 1 \cdots n$, where n is the size of collection $N$)

If we want to get element $entA_i$ in collection $A$ and related element $entB_j$ in collection $B$, original approach takes two queries, while the pre-join approach only takes one. Thus, without considering other factors, the pre-join approach will only take half of the time used before. For collection $A$ with 1,000,000 documents and collection $B$ with 10,000 documents, tests show that queries with the pre-join approach only take 55% to 67% of the time taken by the original approach. With this method, big data of the BIM can be effectively managed and rebuilt, providing the basis for efficient query. The next section discusses the NLP-based method for BIM applications.

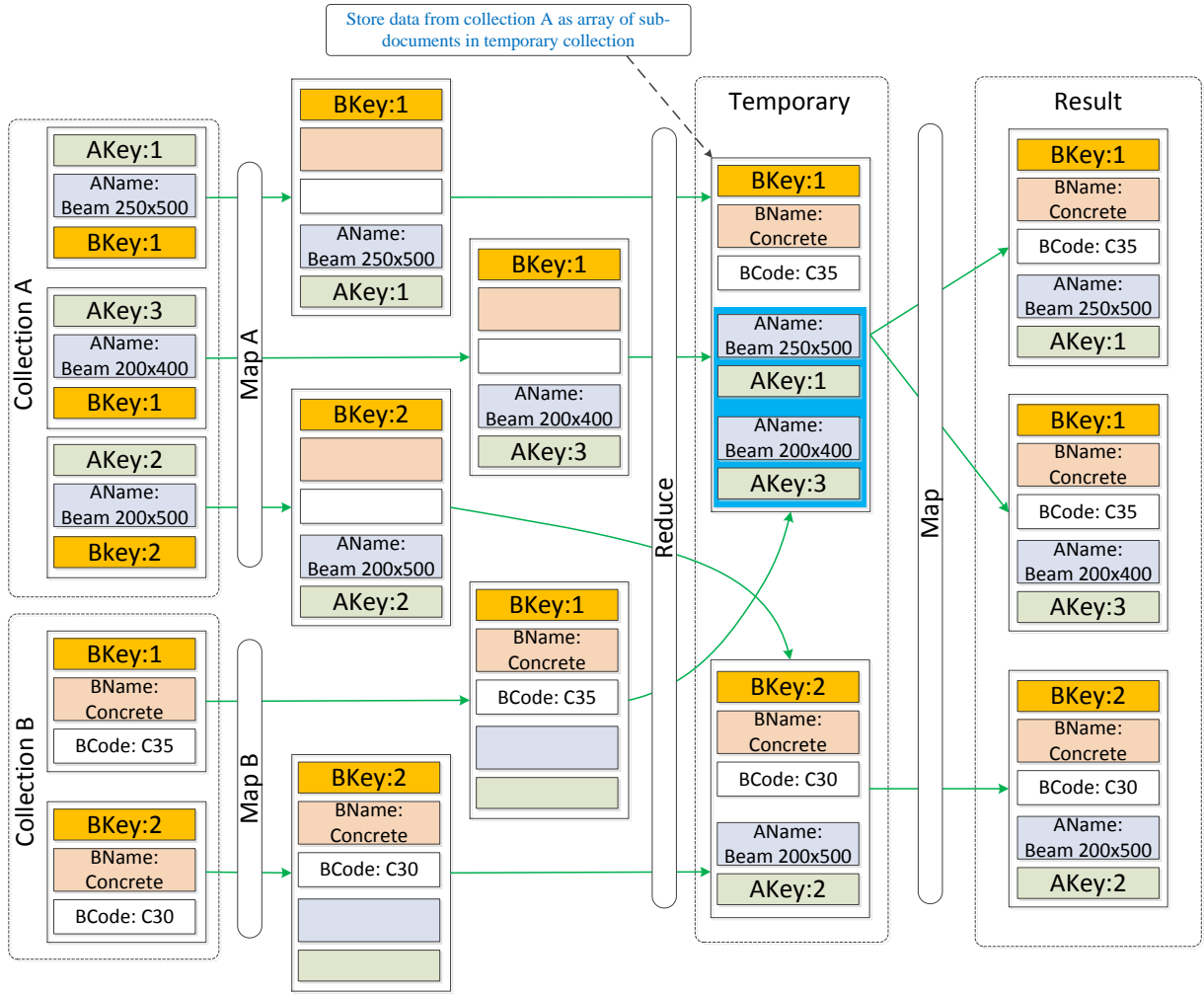

**Figure 3** MapReduce processes to join data in collection B to collection A

## 4 KEYWORD EXTRACTION AND MAPPING

### 4.1 Main process

Understanding the user intention should be first addressed by NLP. Generally, NLP involves the following processes (CCSI, 2012): signal processing, tokenization, syntactic analysis, semantic analysis, and pragmatics. With syntactic tagging and grammar analysis, relations between different segments in a sentence can be determined.

After word segmentation and syntactic parsing, the following questions should be further addressed: (1) Which words represent the main object (like beam, column, etc.) that the users pay attention to? This kind of words can be taken as keywords. (2) Does any constraint related to the keywords exist? It is crucial of this step is to determine which words are keywords, and find their related constraints. The relationships between these keywords and the IFC-based data model should be established. This is domain-dependent, which was not addressed by current NLP tools. Therefore, with keyword extraction and mapping, the foundation for the understanding of user intention and retrieving BIM data can be laid. In the proposed approach, keyword extraction and mapping consists of five steps (Figure 4).

1. Tokenization or word segmentation: refers to the segmentation of user input into words; different NLP algorithms such as compression-based algorithm (Teahan et al., 2000), n-gram model, maximum matching and hidden markov model will be used when processing text in different languages.

2. Tagging: labeling each word of the sentence as noun, verb, and adjective, etc., based on statistical data or other methods.

3. Parsing: to obtain the relationship between different segments of the sentence to form as syntactic structure based on part-of-speech tagging and analysis of the syntactic structure of the sentence.

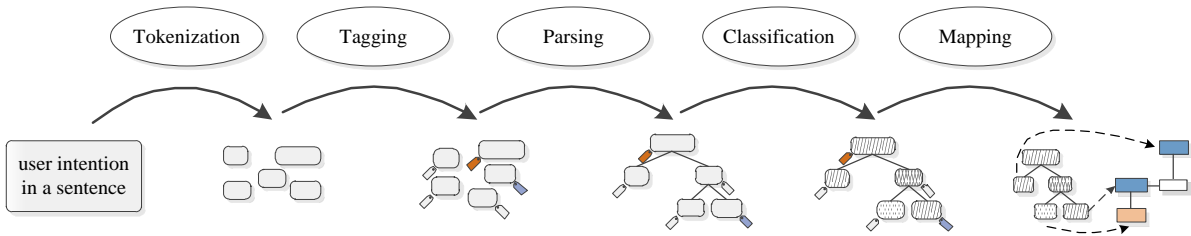

**Figure 4** Steps of keyword extraction and mapping

4. Classification: based on its part of speech, each word will be sorted into different collections, which helps in determining whether a word should map to an entity or its attribute. Herein, these words are exactly "keywords".

5. Mapping: each keyword is mapped to an entity or its attribute in accordance with the relationships between concepts in IFD and entities in an IFC data model.

As the essential parts of NLP, tokenization and parsing for different languages were deeply discussed (Chiarcos et al., 2012; Wang et al., 2014), and numerous NLP tools (such as NLTK and Stanford parser) were already developed. In the current research, among the many NLP tools proposed, the Stanford parser (De Marneffe et al., 2006) was selected to parse the words into a syntactic tree structure. The detail of this process is further discussed in the following part.

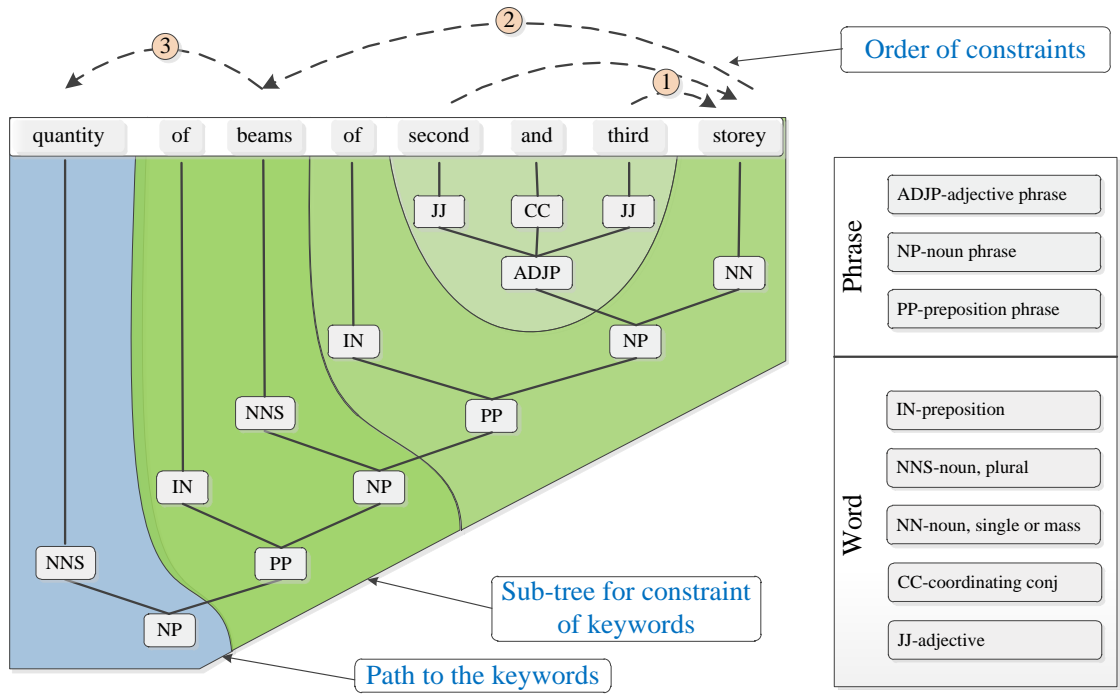

**Figure 5** Tags of segments and their relations in a sentence

## 4.2 Keyword extraction based on NLP

Consider the natural sentence "quantity of beams of second and third storey" as an example. Following the above-mentioned steps, the user intention in a sentence processed by NLP tools will be parsed to a probabilistic context-free grammar structure similar to Figure 5. The sentence is split into words, and all the words are tagged by Penn Treebank POS tag set (Bies et al., 1995). The final syntactic structure is shown as a tree, where ADJP, NP, and PP represent adjective, noun and preposition phrase respectively, and IN, NNS, NN, CC, and JJ stand for preposition, noun (plural), noun (single or mass), coordinating conjunction, and adjective, respectively.

From the root of the tree, a path passes through all nodes only tagged as NP/NN/NNS to the leaf (like left blue part of Figure 5), which is considered the keyword that represents what concerns the user. For each sub-tree whose root is tagged as noun (or other noun phrase), the path to the leaf with all nodes tagged as noun or its equivalent tag determines the keyword of this sub-tree, while the child node tagged as adjective or preposition phrase performs as a constraint for the keyword node that share the same parent with it (middle green part of Figure 5). The order of constraints (top of Figure 5) for different keywords can be determined through recursive analysis of the sub-tree of the structure, which is the inversion of the analysis order. Based on the extracted keywords, it is clear that the key concepts are "quantity", "beam", and "storey". Since adjective and preposition phrases provide more information about the keywords, with the order of constraints, it can be determined that what the user wants to know is the quantity constraint to beams, which were contained in a storey whose name is equal to second or third. These grammatical relationships are important in the subsequent data retrieval and representation.

However, two adjacent leaf nodes with the same parent node may both be tagged as NN in some sentences, for instance "construction progress of the check-in zone" (Figure 6). In this situation, a type dependency analysis based on an NLP toolkit (like Stanford parser) is required, and the dependency between "construction" and "progress" in the sentence is expressed as *NN (progress-2, construction-1)*, which means that the keyword should be "progress" and should be constrained by "construction".

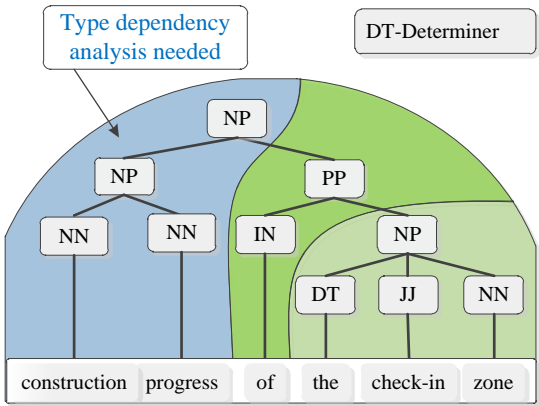

**Figure 6** Tree structure for a sentence with adjacent leaf nodes tagged as NN

Finally, considering the above-mentioned situations, main steps for extracting keywords from a tree structure of a sentence should be as follows: (1) execute a type dependency analysis for the tree structure; (2) traverse the tree for a path connecting a leaf node tagged as NN to the root then return the leaf node; and (3) if two leaf nodes share a same parent, then obtain the keyword based on results of the type dependency analysis.

### 4.3 IFD-based keyword mapping

The IFD Library is an open library, where concepts and terms are defined, semantically described and given a unique identification number. Concepts have two basic types: (1) subjects (or terms): something that is unique and can be distinguished from other things, and (2) characteristics (or properties): concepts that cannot be defined by other concepts. These concepts are classified into six types, namely, behavior, environmental influence, function, measure, property, and unit (NBIMS, 2012). Relations between subjects and characteristics are also classified into different types, namely, "acts upon," "composes," "specializes," "assigns properties," "assigns measures," and "assigns values". All the concepts are assigned with a global unique identifier (GUID) to identify and reuse of the concepts. However, the IFD library is a database for concepts, not for their instances. IFD only defines the names for types of objects and properties that describe them. The relationship between a concept in an IFD library and an entity in an IFC model is established by the GUID assigned to the concept.

Figure 7 shows that, "beam" is defined as a concept and "quantity" is defined as property. "Volume" that specializes in "Physical Quantity" means that "volume" is also a kind of "physical quantity", the same thing goes with "specializes" relationship between subjects. The relationship "assigns properties" defines a property and specifies a characteristic for the concept. Similarly, "acts upon" stands for the relation between an activity and the concept it can apply to, and "composes" specifies the relationship between a concept and its different parts. Meanwhile, entities in an IFC model hold references to concepts defined in the IFD library. As illustrated in Figure 7, each of the entities in IFC model whose type is "IfcBeam" holds a reference to the "Beam" concept, and similar relations are established between "IfcPhysicalQuantity / IfcQuantityVolume" and "Physical Quantity / Volume".

To obtain understandable information that exchanged through IFC, additional information for concepts defined in IFD is needed. A synonym or plural form of a name of an entity, as well as the name of an entity in different languages can be correctly understood, as long as the correct GUID is provided by the IFD library. For example, based on analysis of quantity takeoff and construction management standards in China, the prototype of IFD library previously developed using Chinese and English (Zhang et al., 2012) was extended in the following two aspects to support keyword mapping:

1. Synonym expansion: Terms with the same meaning were related to the same concept by object relationships. Thus, the words "girder" and "beam" should all be related to the term "beam" in the construction domain.

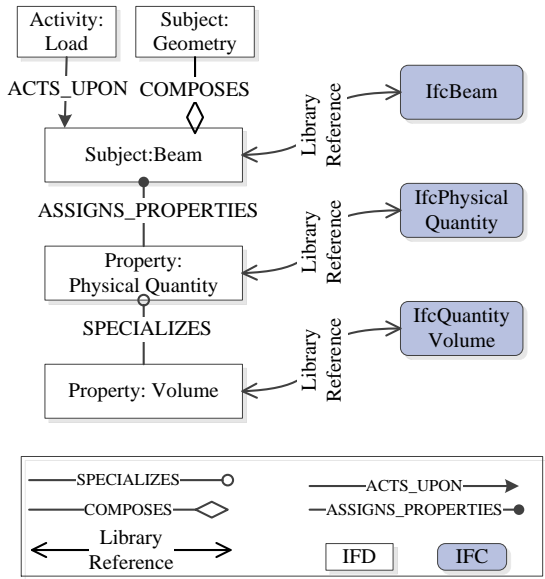

**Figure 7** Structure of an IFD subject and its mapping to IFC

2. Form standardization: Different forms of a term, including abbreviations, different spellings, single/plural form, and even numbers were also related to a "standard" term.

To realize synonym and form standardization, two new relations, namely, "same to" and "alias as" were added to the IFD library. "Same to" stands for the relations between different terms of the same concepts, while "alias as" is used to relate a term to its different forms. 60 synonyms, plural forms of all the concepts were added when extending the IFD library. In addition, different spellings, and abbreviations of a few terms were also added. Finally, an IFD library of more than 700 concepts was established.

Based on the mapping mechanism provided by IFD (BuildingSMART, 2008), extracted keywords can be mapped to IFC entities or attributes. Given that keywords may be in different forms, all the keywords are replaced by their "standard" form, and then, a synonym checking is performed. First, keywords are searched in the IFD library, once their related concepts are found, they can be mapped to IFC entities and attributes according to the relationships between concepts and IFC entities and attributes, thereby supporting data retrieval. In this manner, the words "quantity," "beams," and "storey" will be mapped to "IfcPhysicalQuantity," "IfcBeam," and "IfcBuildingStorey" respectively.

A rule-based approach is also applied to map an abstract term such as "quantity", which means mass when the material is steel and volume when the material is concrete, in order to help in mapping an abstract concept to an IFC entity. Given that some applications export data of quantities into a property set, if no instance of "IfcphysicalQuantity" occurs, then "quantity" should be mapped to "IfcProperty".

Until now, keywords of the sentence that represents the intention of a user can be extracted. Considering the relations between keywords and IFC entities, the IFC-based storage of the information that the user is concerned with is determined by the computer. The next section discusses how to retrieve the information from an IFC-based data model based on the analysis results of this section.

## 5 DATA RETRIEVAL AND REPRESENTATION

### 5.1 Relation finding and constraint transforming for data retrieval

The relations between keywords and IFC entities are established through the method proposed in Section 4. Therefore, the key IFC entities that should be retrieved have been determined. However, the relations between these IFC entities and their attributes are still unknown. Thus, the first step of data retrieval is to determine the relations between different IFC entities or attributes. In this research, a graph-based path search method was adopted to find the relations between entities within the IFC data model.

First, an express file converter was developed to convert the whole IFC schema to a graph for entity relation searching (Figure 8). When converting the IFC schema, all entities and defined types were considered as nodes, and attributes and inheritances were considered as edges. The nodes and edges formed a complex graph with a number of entrances and exits for each node. Then, an xgml file for graph storage was generated from the express file for the IFC schema.

With all relations described in the graph, providing some entities, a path can be found based on the path search algorithm such as Dijkstra. Considering the sentence in Figure 5 as an example, the path that connected "IfcBuildingStorey," "IfcBeam," and "IfcProperty" was found as the bottom of Figure 8.

The path found defines the relations between different entities in the IFC model. To extract the right data from the database, constraints and their logical relations should be considered.

Constraints that define the information "A belongs to B" have been already expressed in the found path, for example, the IfcRelContainedInSpatialStruture that connects IfcBuildingStorey and IfcElement presents the containing relationship of "beams of the second storey." Thus, the value constraints of properties and the logical relationships between different constraints should be considered.

The value constraint of a property in a short sentence always appears like "concrete beam," "rectangular steel column," or "second storey." ($\cup W_c | W_N$) can be used to define the value-constraint of a property of an entity, where $W_N$ is the keyword that related to a subject in an IFD library, and $\cup W_c$ are the words used to specify the values of different properties of the subject. For a value-constraint $W_c$ of a property, the concept that $W_c$ can define or specify in the characteristics of the subject can be searched from the IFD library. For example, in the term "concrete beam," "concrete" can be searched in the properties of the "beam" concept, and can be found in the "material" property. Thus, "concrete" can be considered as a value constraint for the property "material" of "beam". If none is found in properties of the concept, the "name" property can be used as default.

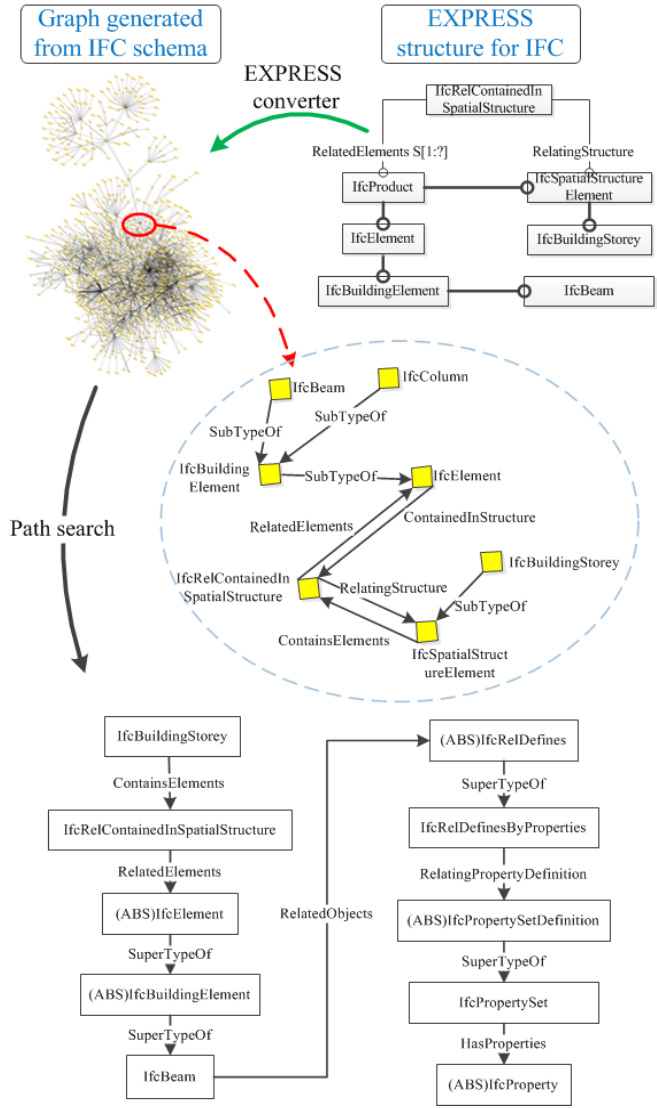

**Figure 8** Data retrieval path found in a graph converted from an IFC schema

To determine the logical relations between different constraints, the basic logical operations in Table 1 can be considered as an example. The basic principles for transforming these logical relationships are as follows: if the word "and" (or "or") connects (1) two noun phrase nodes with their keywords related to the same concept in IFD,

"and" (or "or") should be transformed to the element set union (or intersection). For instance, in "quantity of beams and columns," the data results for "beams" and "columns" retrieved from the database should be united and then used to calculate the quantity; (2) two adjunctive nodes, it should be transformed to logical AND/OR. Thus, the phrase "second or third storey" should be transformed to "instance.Name='second' OR instance.Name='third'", where instance is the instance of IfcBuildingStorey; and (3) other conditions, retrieving, analyzing, and representing the data separately will be effective.

**Table 1**
Operations for "and" and "or" under different conditions

| conditions | word | And | Or |
|---|---|---|---|
| Connecting two adjunctive nodes | | Boolean operation AND | Boolean operation OR |
| Connecting two noun phrases with the same keywords | | Set operation INTERSECTION | Set operation UNION |

With property constraints and logical relations clarified, different logical operations derived from the constraints and relations should be carefully combined. The logical operations from a leaf node of the tree structure should be combined first, then to its parent node and recursively to the root, following the order of constraints implied in the syntactic structure of the sentence.

With the relations between different IFC entities and attributes found and the logical operations and constraints defined, data retrieval can be implemented on different IFC-based databases. In this research, MongoDB which provides a C# API for data retrieval was selected as the data storage platform. Thus, data query function calls can be dynamically generated and invoked for data retrieval. Checking whether a pre-join operation is necessary should be conducted when executing data retrieval functions. If this operation is required, the operation will be executed first based on the MapReduce framework of MongoDB.

The following processes or analyses can be implemented based on the MapReduce framework when retrieving data from the database. They will be automatically executed parallel among different clusters by the MongoDB server:

1. Classification: All the data can be classified into different groups based on the different values of the same attributes. For example, type of a beam is stored in the attribute "ObjectType" of IfcBeam for IFC files exported from Autodesk Revit, and beams with different "ObjectType" are grouped into different groups.

2. Summarization: Once the data have been classified into different groups, a summary for each group can be created. Thus for the same group, the quantity of the building elements or the total number of the elements can be easily calculated.

Data retrieved from the database should be tagged with keywords so that they can be further classified into different groups by keywords. In the above-mentioned example, "quantity of beams" with different materials is classified by material into "weight" and "volume" groups. Each group can then be classified by the storeys that the beams belong to.

**5.2 Data representation**

Data representation, which significantly affects how users understand the data, encompasses formatting data into tables and tree lists as well as different forms of charts. Based on the application and analysis on BIM data representation in the previous research, different data formats were summarized as follows.

1. Single value: the simplest format, like height of a beam;
2. Array: a collection of objects in a similar format;
3. Tree: a set of objects in a similar format, and part of them can be nested in one of the collections, tree format in IFC includes IfcSite, IfcBuilding and IfcSpace;
4. Net structure: objects that are related to one another, such as IfcProcess in IFC;
5. Data that are related to geometries.

To help user obtain a better view of the retrieved data and summarized results, representation methods are listed in Table 2.

**Table 2**
Representation methods for different data formats

| *Data format* | *Representation* | *Methods* |
|---|---|---|
| Single value | Plain text | Value can be highlighted by color or bold font if it has a unit |
| Array | Charts and tables | 1. 1D and 2D arrays that contain string values can be visualized as column and bar charts<br>2. Other 2D and higher-dimensional arrays should be represented as charts with multiple series.<br>3. Time-based data should take time as the X-axis.<br>4. Details of array can be listed in a table for reference. |
| Tree | Grouped charts and tree list | 1. Data in a tree format may use charts in different details.<br>2. Tree list is a suitable way for details. |
| Net structure | Net graph | 1. Best visualization for the net data format is a net graph with labeling.<br>2. Schedule data should be visualized as Gantt chart or timeline.<br>3. Details of the data can be presented in a table or tree list. |
| Data related to geometries | Charts and tables and 3D Geometries | 1. Time-dependent shapes or sequence-related data should be visualized as animation based on a 4D model, with other geometries being invisible.<br>2. Geometries that other data are related to should be colored based on the data they are related to, for example, all shapes of the beams with the same material should be displayed with the same color. And other geometries should be hidden.<br>3. Data can be represented based on the above-mentioned methods. |

## 6 CASE STUDY OF THE TERMINAL OF KUNMING CHANGSHUI AIRPORT

A prototype application called Intelli-BIM was developed based on the above-mentioned methods. The application was tested on the data collected during the construction of the terminal of Kunming Airport.

The terminal of Kunming Airport is located at the northeast of Kunming, the capital city of Yunnan Province in China. It is the hub airport that connects Southeast Asia and South Asia and even Europe. With a total area of 548,300 m$^2$ and a total investment of 6 billion US dollars, the terminal is currently the largest single building in China. The main structure of the terminal consists of seven zones (Zones A, B, C, E, F, G, and H) for a typical floor, with three floors underground and four floors above ground.

The detailed modeling and application process is demonstrated below and in Figure 9.

1. The design BIM was established by Autodesk Revit 2012, including models for architecture/structure, machine, electric, and plumbing (MEP). The design BIM was organized based on the aggregation of spaces and partial spaces as different zones such as zone A, zone B, and check-in zone, and was organized as different systems such as piping and cold water systems based on their functions for building service. Geometric representations of all building elements and hierarchical spatial structure were also provided.

12 Lin et al.

information of different tasks were provided in the Microsoft Project file as text note, and the other construction information including different render styles for activities was modeled in the construction management system. With this approach, the construction BIM for the whole project was established. Part of the model for the MEP integrated unit installation is shown in Figure 9.

3. Quantity takeoff data were modeled by using the quantity bill method (Ma et al., 2011). Different types of quantity such as mass for trusses, length for pipes and cable trays, and number for cameras and water cannons were integrated into the construction BIM.

4. All the construction BIM data were saved in the MongoDB-based cloud database. The applications for intelligent data retrieval and representation were implemented on the Intelli-BIM system.

The cloud consists of three clusters, and each cluster has three servers: one configure server and two data servers. The design BIM of each typical floor consists of approximately 150,000 objects, and the total file size is nearly 1.5 GB. The whole building consists of 1,100,000 objects, and the size of the files is 10 GB in total. With the integrated schedule and quantity takeoff data as well as the associated pictures and documents, the final BIM is approximately 50 GB.

Most of the BIM data were exported and stored in IFC format. The whole building consists of 7 storeys, which were serialized as instances of IfcBuildingStorey with different names and elevations, the storey numbers were stored in the "Name" attribute. Each of the storeys was composed of a few zones. In this research, zones were stored as IfcSpace. The "Name" attribute of IfcSpace was used to distinguish different zones, such as zone A, zone B, and check-in zone, etc. Building elements were related to the zone that contains them by IfcRelContainedInSpatial-Structure.

Given that all BIM data from the design and construction phases were integrated into Intelli-BIM, once user intention described in a natural language sentence was input into the application, user concerned data were extracted, analyzed and represented in different combination forms based on the above-mentioned approach. For example, the sentence "construction progress of the check-in zone" is processed by NLP for keyword extraction and mapping through IFD, providing information on how to retrieve data from the cloud database, and finally, a multi-aspect information representation based on windows form controls provided by Microsoft will be displayed. The feedback of the query is shown in Figure 10 and explained in detail below:

1. The feedback data were represented in a diagram for resources, a detailed list view for the construction schedule, an animation for the construction process simulation and a timeline dashboard for the summary of important tasks during the construction.

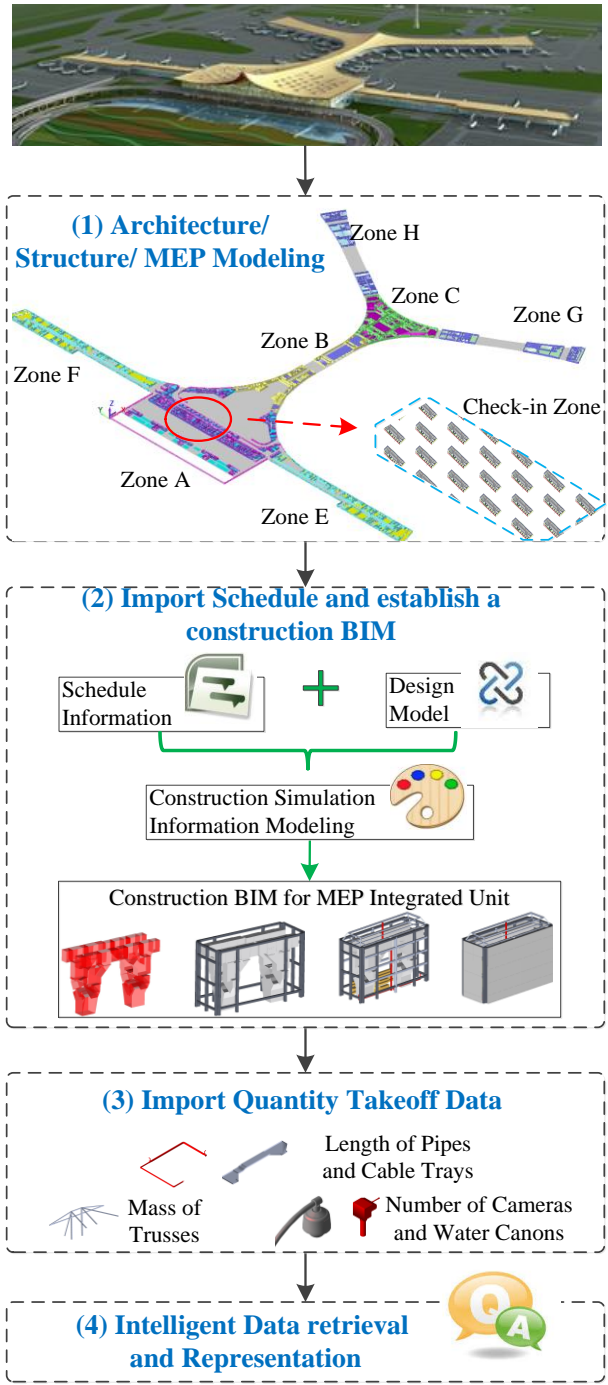

**Figure 9** Modeling and test process

2. By integrating the design BIM exported as IFC files, planned schedule based on Microsoft Project and construction simulation information, a construction BIM was established in the construction management system (Hu and Zhang, 2011) based on an IFC-based graphic information model (Zhang et al., 2014). The contractor

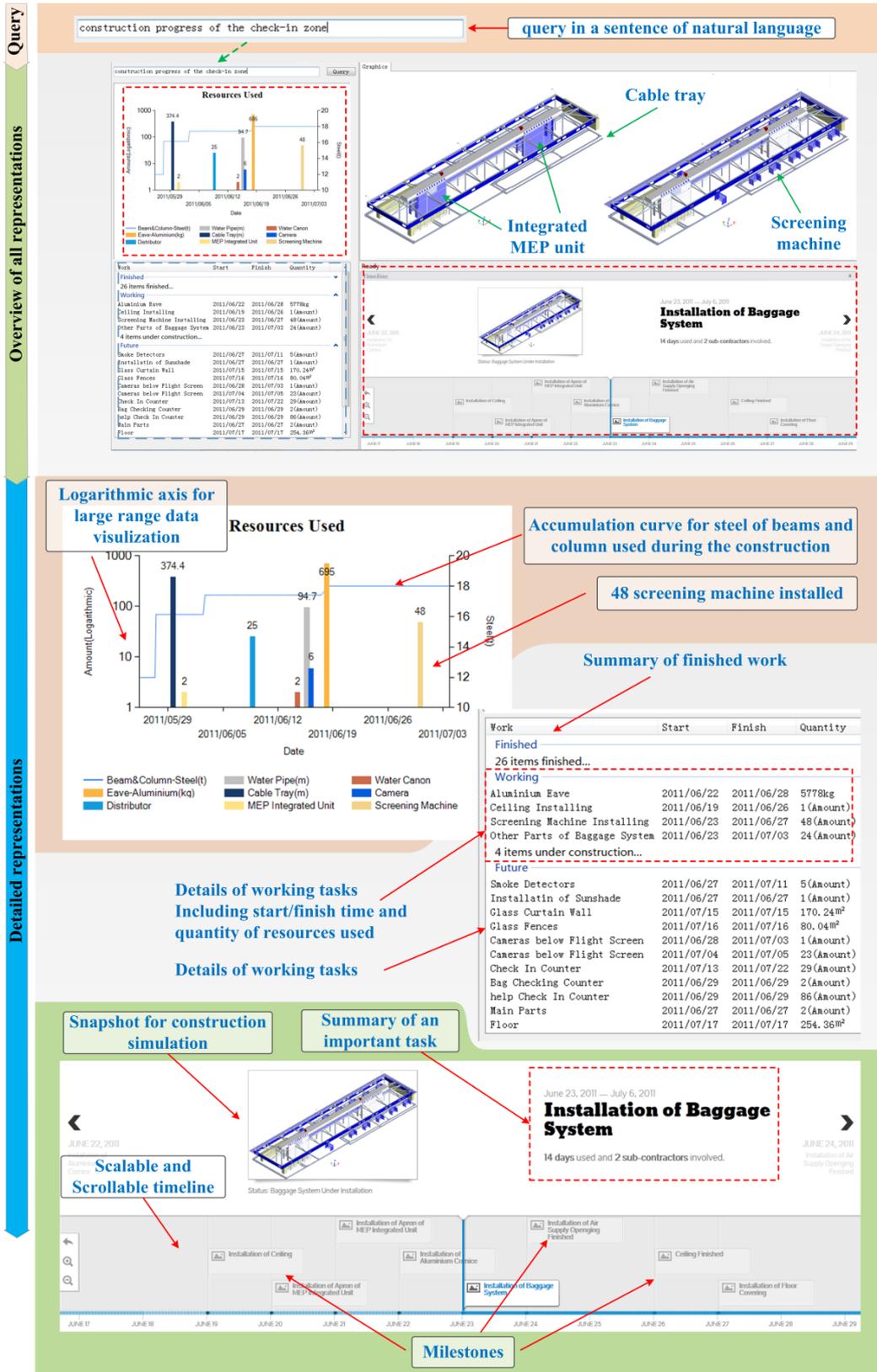

**Figure 10** Application result of intelligent data retrieval and representation

2. For resource usage calculated by aggregating the quantities of building elements related to the tasks, a chart was utilized for result representation. Different types of resources, as previously modeled, were shown in the chart. For resources used in a short time span, a colored bar was chosen, and for long duration resource usage a line strip was adopted. To obtain a feasible representation of different data of different types distributed in a large range, double Y-axis was adopted and configured (i.e., logarithmic and common axes were used, and minimum and maximum values were selected). Thus, users could easily identify how many types of resources were used and their corresponding quantities.

3. A collapsible list view was utilized for details of construction tasks. Finished works were collapsed with a total finished amount as default, and working tasks were displayed with start/finish time as well as resource quantities of the elements related to the task. Since quantities of different resources were displayed in a single column, resource types were not specified in the column header. Works planned in the future were also represented similar to working tasks. Obtaining an overall view of tasks involved in the construction of the check-in zone was flexible for users.

4. For representation of the construction progress, a timeline-based dashboard was proposed. The form of the representation was composed of three parts: a timeline view showed important tasks (such as milestones) as a flag with the task name; a snapshot of the selected task and a title for the picture were provided; and finally, a summary was given including the name, start time, finish time, and duration of the task. Furthermore, related contractors and resources used were presented in the dashboard by the above-mentioned data analysis supports. With the timeline-based dashboard, users could obtain a full view of the construction progress, resulting in an easy and efficient construction management. Such a method could be easily extended to other phases of the project.

5. In summary, with the combination of animations, charts, list views and a timeline-based dashboard, users can easily obtain the whole information for the construction of the check-in zone, proving that: (1) the cloud-based database is suitable for data storage and the pre-join of the data based on the MapReduce framework works fine; (2) the user intention in a natural langue sentence is properly processed, and data retrieval and representation are correctly implemented; and (3) a multi-aspect view that integrates different representations of the results provides users abundant feedback on what they want.

More queries, such as "quantity of beams of second and third storey"(Lin et al., 2013), and "quantity of steel columns of the check-in zone", were tested in the system, the results were shown and displayed as expected. Query time of these queries ranges from 1.5s to 3.5s, keywords extraction and mapping takes about 0.5s, while data retrieval and representation takes the left part of the time. Investigation of each step of the approach shows that: 1) keywords extraction is based on traversal of the syntactic tree provided by the NLP tools, results are mainly limited by these tools; 2) text searching in the IFD library was adopted for keywords mapping, since total number of the concepts is less than 800, it is inevitable that related concepts cannot be found for some keywords, thus resulting in a query failure; 3) data retrieval mainly depends on the relationship finding based on the graph generated from IFC schema. Different time is needed for different number of keywords; 4) only simple sentences without verbs and pronouns were considered, it is inadequate for complex sentences currently.

## 7 CONCLUSION AND FUTURE WORKS

This study explores how a cloud can help in handling the large amount of the BIM data, and how natural-language-based data retrieval and representation can facilitate the request for information of non-experts, thus enhancing the value of BIM. First, to support processing of a huge BIM and to obtain a fast query, a cloud based on MongoDB was established for distributed data storage and the MapReduce framework was adopted to automatically pre-join of two collections. Second, the concepts "keyword" and "constraint" were proposed to capture the key user-concerned objects. An NLP-based method was incorporated in the keyword extraction from the user intention in natural language. Third, an improved method for mapping keywords to IFC entities or attributes based on IFD was devised. Based on the graph generated from the IFC schema, relations between different IFC entities or attributes can be determined. With the evaluation of the constraints of the properties, user-concerned data can be retrieved from the BIM database. Finally, the data were classified, summarized, and then represented based on their formats, such as tables, charts, animations or a combination of these formats. The practical application results in construction management illustrated that with semantic understanding of the user intention in natural language, user-concerned data will be automatically retrieved, analyzed, and represented in proper forms. This process significantly benefits corporations without requiring extremely technological users, thus facilitating BIM application and enhancing the value of BIM.

The presented approach may be adopted in other applications such as facility management, cost management to simplify human-computer interaction. However, the proposed approach does have the following limitations. 1) The results of data representation are not feasible for custom configuration at present. Further works can be done to achieve an flexible information representation. 2) Currently, only simple sentences are supported. Complex sentences containing verbs, sentence with operators (such as *, /), and sentence requiring complex calculation are not supported yet. Further improvement should be made to support mapping verbs and operations to calculation functions properly. 3) Since the IFD library is used for mapping keywords to IFC entities or attributes, the IFD library should contain enough concepts in the AEC domain and be continuously enriched. 4) It is fast to obtain the data in the database with cloud BIM, the bottleneck lays in the keyword extraction and mapping process of the user intention in natural language. Therefore, further improvements like paralleling natural language processing among different cluster and caching frequently used queries should be explored. 5) There is still no enough space on a mobile device for data representation methods presented in this study. Better representation methods are required. 6) Features like full-text search of unstructured documents such as dxf, pdf files should be integrated to take full advantage of BIM.

## ACKNOWLEDGMENTS

The authors would like to thank all the reviewers for their comments and suggestions. This research was supported by the National High Technology Research and Development Program(863 Program) of China (No.2013AA041307), the National Natural Science Foundation of China (No.51278274), and the Tsinghua University-Glodon Joint Research Center for Building Information Model (RCBIM).